\documentclass[lettersize,journal]{IEEEtran}
\usepackage{amsmath,amsfonts}
\usepackage{xcolor}
\usepackage{algorithmic}
\usepackage{algorithm}
\usepackage{array}
\usepackage[caption=false,font=normalsize,labelfont=sf,textfont=sf]{subfig}
\usepackage{textcomp}
\usepackage{stfloats}
\usepackage{url}
\usepackage{verbatim}
\usepackage{graphicx}
\usepackage{cite}

\newcommand {\Define} {\stackrel {\Delta} {=}  }
\newcommand{\mya}{\mathrel{\overset{\makebox[0pt]{{\tiny(a)}}}{=}}}

\newtheorem{theorem}{Theorem}

\begin{document}

%\title{Linear Precoding in Zak-OTFS Systems to Enable Low Complexity Receiver Equalization}
\title{Precoded Zak-OTFS for Per-Carrier Equalization}

\author{Saif Khan Mohammed, Amit Kumar Pathak, Muhammad Ubadah,  Ronny Hadani, Ananthanarayanan Chockalingam and Robert Calderbank
        % <-this % stops a space
\thanks{S. K. Mohammed, A. K. Pathak and M. Ubadah are with the Department of Electrical Engineering, Indian Institute of Technology Delhi, India (saifkmohammed@gmail.com, eez188171@ee.iitd.ac.in, eez198356@ee.iitd.ac.in). R. Hadani is with Department of Mathematics, University of Texas at Austin, TX, USA (hadani@math.utexas.edu). A. Chockalingam is with the Dept. of Electrical Communication Engineering at Indian Institute of Science Bangalore (achockal@iisc.ac.in). R. Calderbank is with the Dept. of Electrical and Computer Engineering, Duke University, USA (robert.calderbank@duke.edu).}

}% <-this % stops a space
%\thanks{Manuscript received April 19, 2021; revised August 16, 2021.}}



\maketitle

\vspace{-10mm} 

\begin{abstract}
In Zak-OTFS (orthogonal time frequency space) modulation the carrier waveform is a pulse in the delay-Doppler (DD) domain, formally a quasi-periodic localized function with specific periods along delay and Doppler. When the channel delay spread is less than the delay period, and the channel Doppler spread is less than the Doppler period, the response to a single Zak-OTFS carrier provides an image of the scattering environment and can be used to predict the effective channel at all other carriers. The image of the scattering environment changes slowly, making it possible to employ precoding at the transmitter. Precoding techniques were developed more than thirty years ago for wireline modem channels (V.34 standard) defined by linear convolution where a pulse in the time domain (TD) is used to probe the one-dimensional partial response channel. The action of a doubly spread channel on Zak-OTFS modulation determines a two-dimensional partial response channel defined by twisted convolution, and we develop a novel precoding technique for this channel. The proposed precoder leads to separate equalization of each DD carrier which has significantly lower complexity than joint equalization of all carriers. Further, the effective precoded channel results in non-interfering DD carriers which significantly reduces the overhead of guard carriers separating data and pilot carriers, which improves the spectral efficiency significantly.
\end{abstract}


\section{Introduction}
Precoding in delay-only channels (e.g. wireline modem channels as in V.34 standard \cite{V341994}) is known to mitigate the effect of inter symbol interference (ISI) and thereby reduce equalization complexity  \cite{Forney98, Palomar2006, Laroia93}. For the doubly-spread channels envisaged in sixth generation (6G) wireless systems \cite{IMT2030}, recently introduced Zak-OTFS modulation achieves significantly better performance than CP-OFDM in high Doppler spread channels \cite{zakotfs1, zakotfs2, otfsbook}. In Zak-OTFS modulation, information symbols are carried  by pulses in the delay-Doppler (DD) domain. Zak-OTFS achieves robustness to high delay and Doppler channel spread by embracing interference between
DD carriers which therefore requires joint equalization of all DD carriers which has high complexity. 
Joint equalization is less of a challenge on the uplink, where the base station (BS) is the receiver, than on the downlink, where the signal processing capability of the user terminal is more limited.
%While joint equalization may be feasible in the uplink where the base station (BS) is the receiver, it is quite challenging in the downlink where the user terminal's receiver usually has limited signal processing capability. 

For delay-only channels, the time-domain (TD) output is given by linear convolution of the TD input with the channel impulse response (CIR). Knowledge of CIR at the transmitter can be used to convolve the information signal with a pre-filter to mitigate ISI thereby reducing equalization complexity \cite{Laroia93}. In doubly-spread channels,
with Zak-OTFS modulation, the input-output (I/O) relation is very simple, the discrete DD domain output signal is given by \emph{twisted convolution} of the discrete DD domain input signal with the effective DD channel filter. Motivated by the pre-filtering idea for delay-only ISI channels, for doubly-spread channels, this paper proposes precoded Zak-OTFS where the transmit DD domain signal is given by twisted convolution of a DD pre-filter with the DD information signal. The pre-filter is designed in such a way that the precoded I/O relation is almost ideal, i.e., DD carriers do not interfere with each other (see Sections \ref{sec3} and \ref{secfilterdesign}). 
%This enables low-complexity per-carrier equalization at the receiver.
The proposed pre-filtering/precoding has several advantages as follows:
\begin{itemize}
\item \emph{Separate} equalization of each DD carrier which has significantly \emph{lower complexity} than joint equalization in conventional Zak-OTFS with no transmitter precoding.
\item Since DD carriers do not interfere in the proposed precoded Zak-OTFS, it suffices to use only a single pilot DD carrier without the need for any guard DD carriers around it. This implies a significantly \emph{lower channel estimation overhead} and therefore a \emph{higher effective spectral efficiency} compared to conventional Zak-OTFS where the pilot carrier is surrounded by guard carriers \cite{twobytwopaper}.
\item Since the pilot and data carriers do not interfere, the required pilot power is significantly smaller for precoded Zak-OTFS which implies better energy efficiency.
\end{itemize}
Numerical results in Section \ref{simsec} for the standardized Vehicular-A channel confirm that the optimal pilot power to data power ratio is about $10$ dB smaller for precoded Zak-OTFS. It is also observed that the effective spectral efficiency of precoded Zak-OTFS is significantly higher than that of conventional Zak-OTFS even for a very high channel Doppler spread of $10$ KHz. Also, precoding does not change the peak-to-average-power-ratio (PAPR) of the transmitted Zak-OTFS signal.

Multi-carrier (MC) OTFS introduced in \cite{Hadani2017} is different
from Zak-OTFS. The Zak-OTFS I/O relation
is predictable, i.e., channel response to a single Zak-OTFS pilot carrier
can be used to accurately predict the response to other Zak-
OTFS data carriers, which is not true for MC-OTFS \cite{zakotfs2, otfsbook}. The
I/O relation can be acquired with low overhead in Zak-OTFS,
but is challenging for MC-OTFS and due to which its relative
performance degrades in high Doppler spread channels \cite{zakotfs2}.
%DD domain precoding of MC-OTFS for equalizer complexity reduction has been proposed in \cite{SLi2023}. The work in \cite{SLi2023} however assumes channel knowledge at the transmitter, which is challenging due to the non-predictable I/O relation in MC-OTFS.

Due to the predictable I/O relation of Zak-OTFS it is simple to estimate the uplink channel at the BS from the received channel response to a single Zak-OTFS pilot carrier transmitted by the user in the uplink \cite{zakotfs2, otfsbook}. Knowledge of the downlink I/O relation is then achieved by exploiting DD domain reciprocity \cite{Fhlawatsch2011}.
%Further, DD representation of the physical channel changes slowly compared to its time-frequency (TF) representation which can enable feedback of the estimated downlink channel from the user to the BS.
Therefore, in this seminal paper, we assume perfect knowledge of the effective DD domain channel at the transmitter and our focus is on the novel twisted convolution based pre-filtering idea. Also, to the best of our knowledge, there is no work so far on twisted convolution based precoding in Zak-OTFS based systems for equalizer complexity reduction, and this paper is the first to do so.

\section{System model}
\label{secsysmodel}
Let $x(t)$ denote the transmitted time-domain (TD) signal having approximate time-duration $T$ and bandwidth $B$. Let $h_{\mbox{\scriptsize{phy}}}(\tau, \nu)$ denote the
DD spreading function of the physical channel. The received
TD signal is then given by \cite{Bello}

{\vspace{-4mm}
\small
\begin{eqnarray}
y(t) & \hspace{-3mm}  = & \hspace{-3mm} \iint  h_{\mbox{\scriptsize{phy}}}(\tau, \nu) \, x(t - \tau) \, e^{j 2 \pi \nu (t - \tau)} \, d\tau \, d\nu \, + \, n(t)
\end{eqnarray}\normalsize}where $n(t)$ is AWGN with power spectral density $N_0$. Next, in Sections \ref{sectx} and \ref{secrx}, we briefly review Zak-OTFS transmitter and receiver as described in detail in \cite{zakotfs1, zakotfs2, otfsbook}.

\subsection{Zak-OTFS modulation at transmitter}
\label{sectx}
Let $\tau_p, \nu_p$ ($\tau_p \, \nu_p = 1$) denote respectively the delay and Doppler period parameters of Zak-OTFS modulation. Let $M \Define B \tau_p$ and $N \Define T \nu_p$ be integers. Let $x[k,l]$, $k=0,1,\cdots, M-1$, $l=0,1,\cdots, N-1$ denote the $MN$ transmitted symbols. These are embedded into a discrete DD domain signal  $x_{dd}[k,l]  =  x[k \, \mbox{\small{mod}} \, M, l \, \mbox{\small{mod}} \, N] \, e^{j 2 \pi \lfloor \frac{k}{M} \rfloor \frac{l}{N}}$, which is quasi-periodic, i.e., for all $k,l, n, m \in {\mathbb Z}$

{\vspace{-4mm}
\small
\begin{eqnarray}
x_{dd}[k + nM, l + mN] & = & e^{j 2 \pi \frac{n l}{N}} \, x_{dd}[k,l].
\end{eqnarray}\normalsize}Note that TD realization is defined only for quasi-periodic DD signals. The discrete DD domain signal is then converted to a continuous DD domain signal

{\vspace{-4mm}
\small
\begin{eqnarray}
    x_{dd}(\tau, \nu) & \hspace{-3mm} = & \hspace{-3mm}  \sum\limits_{k=-\infty}^{\infty} \sum\limits_{l=-\infty}^{\infty} x_{dd}[k,l] \, \delta(\tau - k/B) \, \delta(\nu - l/T),
\end{eqnarray}\normalsize}which is quasi-periodic with $\tau_p$ and $\nu_p$ as its delay and Doppler periods, i.e. for all $n,m \in {\mathbb Z}$,
   $x_{dd}(\tau + n \tau_p, \nu + m \nu_p)  =  e^{j 2 \pi n \nu \tau_p} \, x_{dd}(\tau, \nu)$.
To limit the time duration and bandwidth of $x(t)$, $x_{dd}(\tau, \nu)$ is filtered with a pulse shaping filter $w_{tx}(\tau, \nu)$ resulting in

{\vspace{-4mm}
\small
\begin{eqnarray}
   x_{dd}^{w_{tx}}(\tau, \nu) & = & w_{tx}(\tau, \nu) \, *_{\sigma} \, x_{dd}(\tau, \nu),  \nonumber \\
   & & \hspace{-24mm} =  \iint w_{tx}(\tau', \nu') \, x_{dd}(\tau - \tau', \nu - \nu') \, e^{j 2 \pi \nu'(\tau - \tau')} \, d\tau' d\nu',
\end{eqnarray}\normalsize}where $*_{\sigma}$ denotes twisted convolution. The transmitted signal $x(t)$ is the TD realization of $x_{dd}^{w_{tx}}(\tau, \nu)$ and is given by its inverse Zak transform
$x(t)  =  {\mathcal Z}_t^{-1} \left( x_{dd}^{w_{tx}}(\tau, \nu) \right) \, =  \, \sqrt{\tau_p} \int\limits_{0}^{\nu_p} x_{dd}^{w_{tx}} (t, \nu) \, d\nu$.

\subsection{Zak-OTFS receiver and DD domain I/O relation}
\label{secrx}
At the receiver, Zak transform of the received TD signal $y(t)$ gives its DD domain representation $y_{dd}(\tau, \nu)$ \cite{otfsbook}

{\vspace{-4mm}
\small
\begin{eqnarray}
    y_{dd}(\tau, \nu) & \hspace{-3mm} = &  \hspace{-3mm} {\mathcal Z}_t\left( y(t) \right) = \sqrt{\tau_p} \sum\limits_{n \in {\mathbb Z}} y(\tau + n \tau_p) \, e^{-j 2 \pi n \nu \tau_p},
\end{eqnarray}\normalsize}which is then match filtered with $w_{rx}(\tau, \nu)$ resulting in

{\vspace{-4mm}
\small
\begin{eqnarray}
y_{dd}^{w_{rx}}(\tau, \nu) & = & w_{rx}(\tau, \nu) \, *_{\sigma} \, y_{dd}(\tau, \nu).
\end{eqnarray}\normalsize}DD domain sampling then gives

{\vspace{-4mm}
\small
\begin{eqnarray}
\label{eqn999}
    y_{dd}[k,l] & \Define & y_{dd}\left( \tau = \frac{k}{B} \,,\, \nu = \frac{l}{T} \right) \nonumber \\
    & = & h[k,l] *_{\sigma} x_{dd}[k,l] \, + \, n_{dd}[k,l] \nonumber \\
    & & \hspace{-19mm} = \hspace{-2mm} \sum\limits_{k',l' \in {\mathbb Z}} \hspace{-2mm} h[k',l'] \, x_{dd}[k - k', l- l' ] \, e^{j 2 \pi l' \frac{(k - k') }{MN}} \, + \, n_{dd}[k,l],
\end{eqnarray}\normalsize}where the effective DD domain channel filter $h[k,l]$ is

{\vspace{-4mm}
\small
\begin{eqnarray}
    h[k,l] & \Define & h\left( \tau = \frac{k}{B} \,,\, \nu = \frac{l}{T} \right), \nonumber \\
    h(\tau, \nu) & \Define & w_{rx}(\tau, \nu) *_{\sigma} h_{\mbox{\scriptsize{phy}}}(\tau, \nu) *_{\sigma} w_{tx}(\tau, \nu),
\end{eqnarray}\normalsize}and $n_{dd}[k,l], k,l \in {\mathbb Z}$ is the filtered and sampled AWGN.
\section{Proposed DD domain precoding}
\label{sec3}
Let $s[k,l]$, $k=0,1,\cdots, M-1$, $l=0,1,\cdots, N-1$ denote the $MN$ information symbols which are embedded into the quasi-periodic discrete DD domain information signal

{\vspace{-4mm}
\small
\begin{eqnarray}
\label{eqn1111}
     s_{dd}[k,l] & = & e^{j 2 \pi \lfloor \frac{k}{M} \rfloor \, \frac{l}{N} } \, s[k \, \mbox{\scriptsize{mod}} \, M, l \, \mbox{\scriptsize{mod}} \, N].
\end{eqnarray}\normalsize}This information signal is precoded into the transmit DD signal $x_{dd}[k,l]$ by filtering with a precoding filter $a[k,l]$ i.e.

{\vspace{-4mm}
\small
\begin{eqnarray}
\label{eqn1212}
x_{dd}[k,l] = a[k,l] \, *_{\sigma} \, s_{dd}[k,l].
\end{eqnarray}\normalsize}Note that since $s_{dd}[k,l]$ is quasi-periodic and twisted convolution preserves quasi-periodicity \cite{zakotfs1, zakotfs2,otfsbook}, $x_{dd}[k,l]$ is also quasi-periodic. Using this in (\ref{eqn999}) gives

{\vspace{-4mm}
\small
\begin{eqnarray}
\label{eqn1313}
    y_{dd}[k,l] & \hspace{-3mm} = & \hspace{-3mm} h[k,l] *_{\sigma} a[k,l] *_{\sigma} s_{dd}[k,l]   \, + \, n_{dd}[k,l] \nonumber \\
    & = &  h_a[k,l] *_{\sigma} s_{dd}[k,l] \, + \, n_{dd}[k,l],\nonumber \\
    h_a[k,l] & \Define & h[k,l] *_{\sigma} a[k,l],
\end{eqnarray}\normalsize}where the second step follows from the associativity of the twisted convolution operation.\footnote{\footnotesize{In (\ref{eqn1212}) we carry out pre-filtering using twisted convolution. 
%and not linear convolution since the cascade of linear convolution followed by twisted convolution with the channel filter cannot be described simply as an operation between the input information signal $s_{dd}[k,l]$ and some net effective precoded channel filter, and therefore would make it challenging to acquire the precoded I/O relation at the receiver. However, with the proposed twisted convolution pre-filtering
Associativity of the twisted convolution operation ensures that
the DD domain output is simply given by twisted convolution of the DD information signal $s_{dd}[k,l]$ with a net effective precoded channel filter $h_a[k,l]$ which is simply the twisted convolution of the pre-filter $a[k,l]$ and the channel filter $h[k,l]$. The precoded I/O relation is therefore predictable and can be acquired at the receiver with low overhead.}} Let ${\mathcal S}_h    \Define   {\{} \,  (k,l) \, {\vert} \, k_{min} \leq k \leq k_{max} \,,\, -l_{max} \leq l \leq l_{max} \, {\}}$ denote the support set of $h[k,l]$ ($k_{min} \leq 0 \leq k_{max}$, $(k_{max} - k_{min}) < M$, $0 \leq l_{max} < \frac{N}{2}$). We choose the support set of $a[k,l]$ in such a way that the support set of $h_a[k,l]$ is $[ -M/2 \,,\, M/2 - 1 ] \times [-N/2 \,,\, N/2 - 1]$.
Since $h_a[k,l] = h[k,l] *_{\sigma} a[k,l]$, we accordingly choose the support set of $a[k,l]$ to be ${\mathcal S}_a   \Define  {\{}  \, (k,l) \, {\vert} \, -\frac{M}{2} - k_{min} \, \leq \, k \, \leq \, \frac{M}{2} - k_{max} - 1 \,,\, -\frac{N}{2} + l_{max} \, \leq \, l  \, \leq \, \frac{N}{2} - l_{max} - 1 \, {\}}$. With i.i.d. zero mean information/data symbols $s[k,l]$
each having energy $E$, unit energy transmit pulse shaping filter and matched receive filter (i.e. $\iint \vert w_{tx}(\tau, \nu) \vert^2 \, d\tau \, d\nu = \iint \vert w_{rx}(\tau, \nu) \vert^2 \, d\tau \, d\nu = 1$), and a normalized precoding filter, i.e.

{\vspace{-4mm}
\small
\begin{eqnarray}
\label{eqn1616}
    \sum\limits_{(k,l) \in {\mathcal S}_a } \vert a[k,l] \vert^2 & = & 1,
\end{eqnarray}\normalsize}the average received information/data energy on each DD domain carrier/bin is $E$, i.e., ${\mathbb E}\left[ \vert h_a[k,l] *_{\sigma} s_{dd}[k,l]  \vert^2 \right] = E$ (here we assume a normalized channel, where the sum of the mean-squared absolute values of the channel path gains is unity) \cite{zakotfs2, otfsbook}.
For unit energy receive match filter $w_{rx}(\tau, \nu)$, the
variance of $n_{dd}[k,l]$ is $N_0$. The ratio of the received data energy
to the noise variance on each DD carrier is the signal to noise ratio (SNR) $\rho \Define E/N_0$.
We arrange the $MN$ taps of the support set of $h_a[k,l]$ into a vector

{\vspace{-4mm}
\small
\begin{eqnarray}
\label{eqn1717}
    {\bf h_a} & = & \left( h_{a,0} \,,\, h_{a,1} \,,\, \cdots \,,\, h_{a, (MN-1)} \right)^T  \nonumber \\
    h_{a, k+\frac{M}{2} + \left(l + \frac{N}{2}\right) M} & = & h_a[k,l],
\end{eqnarray}\normalsize}$k = -M/2, \cdots, M/2-1$, $l=-N/2, \cdots, N/2 -1 $. From $h_a[k,l] = h[k,l] *_{\sigma} a[k,l]$=$\sum\limits_{k',l'} a[k', l'] h[k - k', l - l'] e^{j 2 \pi k' \frac{(l - l' )}{MN}}$, it follows that

{\vspace{-4mm}
\small
\begin{eqnarray}
\label{eqn1818}
    {\bf h_a} & = & {\bf H} \, {\bf a},
\end{eqnarray}\normalsize}where ${\bf a} \in {\mathbb C}^{K \times 1}$, $K = (k_{min} - k_{max} + M)(N - 2l_{max})$, is the vector of taps of the filter $a[k,l]$

{\vspace{-4mm}
\small
\begin{eqnarray}
    {\bf a} = \left( a_0 \,,\, a_1 \,,\, \cdots, a_{K-1} \right)^T, & & \nonumber \\
    & & \hspace{-60mm} a_{k + \frac{M}{2} + k_{min} + \left(l + \frac{N}{2} - l_{max}\right) (M - k_{max} + k_{min})}  \Define  a[k,l],
\end{eqnarray}\normalsize}$k = -\frac{M}{2} - k_{min}, \cdots, \frac{M}{2} - k_{max}- 1$, $l = -\frac{N}{2} + l_{max}, \cdots, \frac{N}{2} - l_{max}- 1$, and
the matrix ${\bf H} \in {\mathbb C}^{MN \times K}$ with the element in its $\left(k+\frac{M}{2} + \left(l + \frac{N}{2}\right) M \right)$-th row and $k' + \frac{M}{2} + k_{min} + \left(l' + \frac{N}{2} - l_{max}\right) (M - k_{max} + k_{min})$-th column, $k = -M/2, \cdots, M/2-1$, $l=-N/2, \cdots, N/2 -1 $, $k' = -\frac{M}{2} - k_{min}, \cdots, \frac{M}{2} - k_{max}- 1$, $l' = -\frac{N}{2} + l_{max}, \cdots, \frac{N}{2} - l_{max}- 1$, given by

{\vspace{-4mm}
\small
\begin{eqnarray}
    \hspace{-2mm} H_{\left(k+\frac{M}{2} + \left(l + \frac{N}{2}\right) M\right), \left( k' + \frac{M}{2} + k_{min} + \left(l' + \frac{N}{2} - l_{max}\right) (M - k_{max} + k_{min}) \right)} &  & \nonumber \\
    & & \hspace{-91mm} \Define   \begin{cases}
        h[k - k', l - l'] e^{j 2 \pi k' \frac{(l - l' )}{MN}} \,,\, (k - k' , l - l') \in {\mathcal S}_h \\
        0 \,,\, \mbox{\small{otherwise}} \\
    \end{cases}.
\end{eqnarray}\normalsize}

\section{Proposed design of precoding filter}
\label{secfilterdesign}
The precoding filter is designed in such a way that an estimate of the $(k,l)$-th information symbol $s[k,l]$, $k=0,1,\cdots, M-1, l=0,1,\cdots, N-1$, can be acquired directly from the received symbol $y_{dd}[k,l]$ on the $(k,l)$-th DD tap.
In (\ref{eqn1313}), the signal received at the $(k,l)$-th DD tap ($k=0,1,\cdots, M-1, l=0,1,\cdots, N-1$) is

{\vspace{-4mm}
\small
\begin{eqnarray}
\label{eqn18192}
    y_{dd}[k,l] & \hspace{-3mm} = & \hspace{-3mm} \underbrace{h_a[0,0] \, s_{dd}[k,l]}_{\mbox{\scriptsize{useful term}}}  \nonumber \\
    & & \hspace{-18mm}+  \hspace{-4mm} \underbrace{\sum\limits_{(k',l') \ne (0,0)} \hspace{-2mm} h_a[k',l'] s_{dd}[k - k', l - l'] \, e^{j 2 \pi l' \frac{(k - k')}{MN}}}_{\mbox{\scriptsize{Interference term}}} \, + \, n_{dd}[k,l].
\end{eqnarray}\normalsize}The signal to interference and noise ratio is therefore

{\vspace{-4mm}
\small
\begin{eqnarray}
\label{sinreqn1}
    \gamma({\bf a}) & = &  \frac{\vert h_a[0,0] \vert^2}{\frac{1}{\rho} + \sum\limits_{(k,l) \ne (0,0)} \vert h_a[k,l] \vert^2 },
\end{eqnarray}\normalsize}which follows from (\ref{eqn1111}), the fact that summation in the interference term is limited to DD taps $(k,l)$ in the support of $h_a[\cdot, \cdot]$ which is equal to one period along the delay and Doppler axis, the fact that the information symbols $s[k,l]$ are i.i.d. zero mean with variance $E$, and the fact that the variance of $n_{dd}[k,l]$ is $N_0$. 

Let ${\bf h}_i^H \in {\mathbb C}^{1 \times K}$, denote the $i$-th row of ${\bf H}$, $i=0,1,\cdots, (MN -1)$. From (\ref{eqn1717}) and (\ref{eqn1818}) it follows that $h_a[0,0] = {\bf h}_{\frac{M}{2} (N +1)}^H \, {\bf a}$. Therefore, (\ref{sinreqn1}) can be re-written as

{\vspace{-4mm}
\small
\begin{eqnarray}
\label{eqn2323}
\gamma({\bf a}) & = & \frac{\vert {\bf h}_{\frac{M}{2} (N +1)}^H {\bf a} \vert^2}{\frac{1}{\rho} \, + \, \sum\limits_{i=0, i \ne \frac{M}{2}(N+1)}^{MN -1} \vert {\bf h}_i^H {\bf a} \vert ^2 }.
\end{eqnarray}\normalsize}The next theorem gives the optimal ${\bf a}$ which maximizes $\gamma({\bf a})$ subject to (\ref{eqn1616}) (this constraint in (\ref{eqn1616}) is same as ${\bf a}^H {\bf a} = 1$).

\begin{theorem}
\label{thm1}

{\vspace{-4mm}
\small
\begin{eqnarray}
\label{eqn2424}
   {\bf a}_{\mbox{\scriptsize{opt}}} & \Define &  \arg \max_{{\bf a} \vert {\bf a}^H {\bf a} = 1} \, \gamma({\bf a}) \nonumber \\
   & = & \frac{1}{\lambda} \left( \frac{{\bf I}}{\rho} \, + \, \hspace{-4mm}  \sum\limits_{i=0, i \ne \frac{M}{2}(N+1)}^{MN -1} \hspace{-5mm} {\bf h}_i {\bf h}_i^H \right)^{-1} {\bf h}_{\frac{M}{2}(N+1)}
\end{eqnarray}\normalsize}

{\vspace{-4mm}
\small
\begin{eqnarray}
   \lambda & \Define & \left\Vert \left( \frac{{\bf I}}{\rho} \, + \, \hspace{-4mm}  \sum\limits_{i=0, i \ne \frac{M}{2}(N+1)}^{MN -1} \hspace{-5mm} {\bf h}_i {\bf h}_i^H \right)^{-1} {\bf h}_{\frac{M}{2}(N+1)} \right\Vert,
    \end{eqnarray}\normalsize}where $\Vert {\bf g} \Vert \Define \sqrt{{\bf g}^H {\bf g}}$ for any vector ${\bf g}$, and ${\bf I}$ denotes the $K \times K$ identity matrix.
    %The optimal SINR is
    %\begin{eqnarray}
    %\label{sinropteqn}
    %    \mbox{\small{SINR}}_{\mbox{\scriptsize{opt}}} & \Define & \gamma\left(  {\bf a}_{\mbox{\scriptsize{opt}}} \right) \nonumber \\
     %   & & \hspace{-20mm} =  {\bf h}_{\frac{M}{2}(N+1)}^H \left( \frac{{\bf I}}{\rho} \, + \, \hspace{-4mm}  \sum\limits_{i=0, i \ne \frac{M}{2}(N+1)}^{MN -1} \hspace{-5mm} {\bf h}_i {\bf h}_i^H \right)^{-1} {\bf h}_{\frac{M}{2}(N+1)}.
    %\end{eqnarray}
\end{theorem}
\begin{IEEEproof}
Define the $K \times K$ matrix

{\vspace{-4mm}
\small
\begin{eqnarray}
{\bf B} & \Define  & {\bf I}/\rho + \hspace{-6mm} \sum\limits_{i=0, i \ne \frac{M}{2}(N+1)}^{MN -1} \hspace{-7mm} {\bf h}_i {\bf h}_i^H,
\end{eqnarray}\normalsize}For any non-zero ${\bf g} \in {\mathbb C}^{K \times 1}$, it is clear that ${\bf g}^H B {\bf g} = \left( \frac{\Vert {\bf g} \Vert^2}{\rho} + \hspace{-6mm} \sum\limits_{i=0, i \ne \frac{M}{2}(N+1)}^{MN -1} \hspace{-7mm} \vert {\bf g}^H {\bf h}_i \vert^2  \right) > 0$, i.e., ${\bf B}$ is a full rank positive definite matrix \cite{Hornjohnsonbook}. Let ${\bf G} \in {\mathbb C}^{K \times K}$ denote a square-root of ${\bf B}$, i.e., ${\bf B} = {\bf G}^H {\bf G}$. Clearly ${\bf G}$ is also full rank. Re-writing the SINR expression in (\ref{eqn2323}) in terms of ${\bf G}$ gives

{\vspace{-4mm}
\small
\begin{eqnarray}
    \gamma({\bf a}) & = &  \frac{{\bf a}^H {\bf h}_{\frac{M}{2} (N +1)} {\bf h}_{\frac{M}{2} (N +1)}^H {\bf a}}{{\bf a}^H {\bf G}^H {\bf G}  {\bf a}}.
\end{eqnarray}\normalsize}From the expression above it is clear that $\gamma({\bf a})$ is invariant to scaling of ${\bf a}$ by a scalar and therefore the optimum ${\bf a}$ in (\ref{eqn2424}) is obtained from unconstrained optimization and then normalizing ${\bf a}$ so that $\Vert {\bf a} \Vert = 1$. The unconstrained optimization is

{\vspace{-4mm}
\small
\begin{eqnarray}
\label{eqn2727}
    \max_{{\bf a}}  \frac{{\bf a}^H {\bf h}_{\frac{M}{2} (N +1)} {\bf h}_{\frac{M}{2} (N +1)}^H {\bf a}}{{\bf a}^H {\bf G}^H {\bf G}  {\bf a}} & \mya  & \max_{{\bf b}} \frac{{\bf b}^H {\bf F} {\bf b}}{ {\bf b}^H {\bf b}}, \nonumber \\
    & & \hspace{-50mm} {\bf F} \, \Define \, \left({\bf G}^H \right)^{-1} \, {\bf h}_{\frac{M}{2} (N +1)} {\bf h}_{\frac{M}{2} (N +1)}^H {\bf G}^{-1}
\end{eqnarray}\normalsize}where step (a) follows from substituting the optimization variable ${\bf a}$ with ${\bf b} = {\bf G} {\bf a}$ which is equivalent since ${\bf G}$ is invertible. In (\ref{eqn2727}), the optimal ${\bf b}$ is any scalar multiple of the eigen vector corresponding to the largest eigen value of ${\bf F}$ \cite{Hornjohnsonbook}. Since ${\bf F}$ is a rank one matrix, optimal ${\bf b}$ is

{\vspace{-4mm}
\small
\begin{eqnarray}
    {\bf b}_{\mbox{\scriptsize{opt}}} & = & \alpha \left({\bf G}^H \right)^{-1} \, {\bf h}_{\frac{M}{2} (N +1)}
\end{eqnarray}\normalsize}where $\alpha$ is any scalar. Since ${\bf b} = {\bf G} {\bf a}$, the corresponding optimal ${\bf a}$ for the unconstrained optimization in (\ref{eqn2727}) is
$\, {\bf G}^{-1} {\bf b}_{\mbox{\scriptsize{opt}}}  =  \alpha {\bf G}^{-1} \left({\bf G}^H \right)^{-1} \, {\bf h}_{\frac{M}{2} (N +1)}  =  \alpha \, {\bf B}^{-1} {\bf h}_{\frac{M}{2} (N +1)}$. Normalizing this gives the expression for ${\bf a}_{\mbox{\scriptsize{opt}}}$ in (\ref{eqn2424}).
\end{IEEEproof}

\section{Acquisition of the precoded I/O relation and one-tap equalization at the receiver}
\label{sec5}
For the precoding filter given by Theorem \ref{thm1},
it is observed that almost all energy of the effective precoded channel filter $h_a[k,l]$ is \emph{localized at the origin} $(k,l) = (0,0)$ if we operate deep within the crystalline regime (i.e., when the delay and Doppler period are much greater than the spread of $h[k,l]$ along the delay and Doppler axis respectively, equivalently, $(k_{max} - k_{min}) \ll M$ and $2 l_{max} \ll N$).
Next, through numerical examples we illustrate this localized nature of the precoded channel filter. We consider the six path Veh-A channel model \cite{EVAITU} with $h_{\mbox{\scriptsize{phy}}}(\tau, \nu) = \sum\limits_{i=1}^6 h_i \delta(\tau - \tau_i) \delta(\nu - \nu_i)$, where $h_i, \tau_i, \nu_i$ denote the complex gain, delay and Doppler shift of the $i$-th channel path. The power delay profile is given in Table \ref{tab1}. The gain $h_i$ of the $i$-th path is complex Gaussian distributed and ${\mathbb E}[\vert h_i \vert^2]/{\mathbb E}[\vert h_1 \vert^2]$ is the relative power of the $i$-th path w.r.t. the first path (see Table \ref{tab1}). We normalize the path gains so that ${\mathbb E}[ \sum\limits_{i=1}^6 \vert h_i \vert^2] =1$. The Doppler shifts $\nu_i = \nu_{max} \cos(\theta_i)$, where $\theta_i, i=1,2,\cdots, 6$ are i.i.d. uniformly distributed in $[0 \,,\, 2 \pi)$. We consider $\nu_p = 30$ KHz, $\tau_p = 1/\nu_p = 33.33 \, \mu s$, $M = N = 18$. Therefore $B = M \nu_p = 540$ KHz, and $T = N \tau_p = 0.6$ ms. We use the unit energy root raised cosine (RRC) pulse shaping filter at the transmitter \cite{SHDigcomm}, i.e.

\begin{figure}[!t]
\vspace{-1mm}
    \centering
    \includegraphics[width=6.4cm, height=4.1cm]{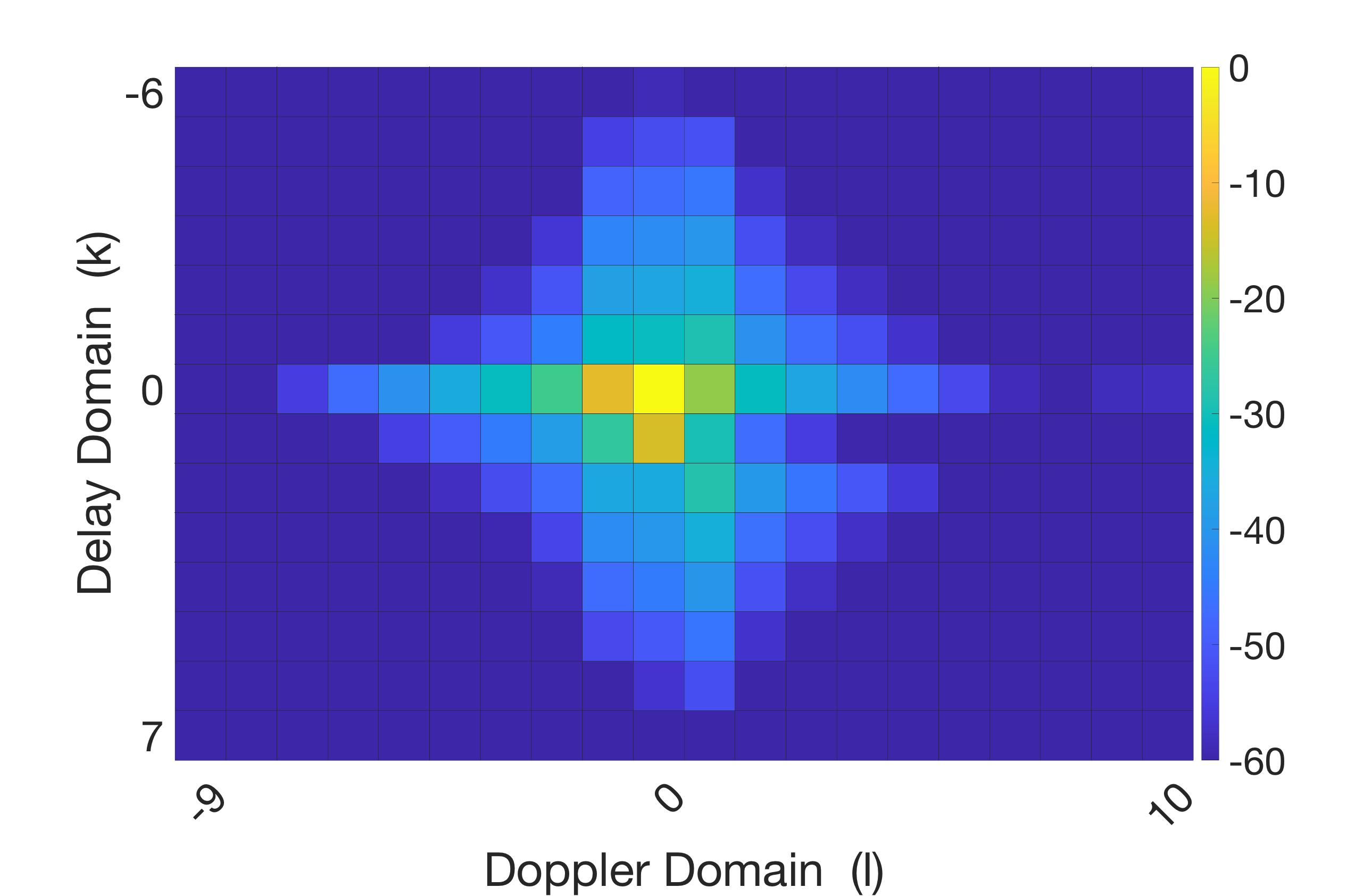}
    \vspace{-2mm}
    \caption{Heat map of $\vert h[k,l]\vert^2$. $\nu_{max} = 1$ KHz.}
    \label{fig1}
    \vspace{-3mm}
\end{figure}

\begin{figure}[!t]
    \centering
    \includegraphics[width=6.4cm, height=4.1cm]{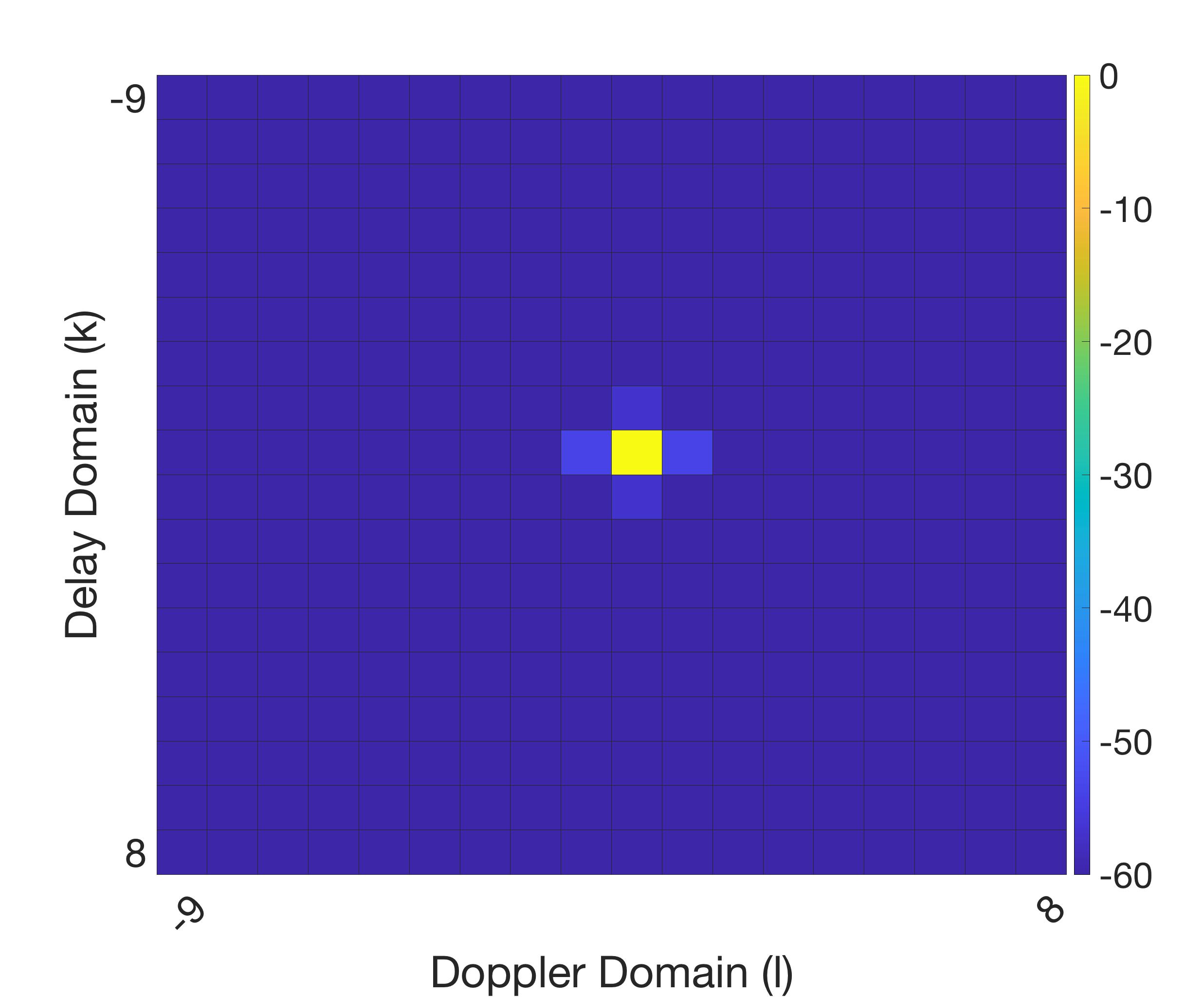}
    \vspace{-3mm}
    \caption{Heat map of $\vert h_a[k,l]\vert^2$ with optimal ${\bf a}$ in (\ref{eqn2424}). $\nu_{max} = 1$ KHz.}
    \label{fig2}
    \vspace{-3mm}
\end{figure}

{\vspace{-4mm}
\small
\begin{eqnarray}
\label{rrcpulse_eqn1}
w_{tx}(\tau,\nu) & = &  \sqrt{BT} \, rrc_{_{\beta_{\tau}}}( B \tau ) \,  rrc_{_{\beta_{\nu}}}( T \nu ), \nonumber \\
rrc_{_{\beta}}(x) &  = &  \frac{\sin(\pi x (1 - \beta)) + 4 \beta x \cos(\pi x (1 + \beta))}{\pi x \left( 1 - (4 \beta x)^2 \right)},
\end{eqnarray}\normalsize}where $\beta_{\tau}$ and $\beta_{\nu}$ are the roll-off factors. Due to time and bandwidth expansion, the actual time-duration and bandwidth of the Zak-OTFS frame is $(1 + \beta_{\nu})T$ and $(1 + \beta_{\tau})B$ respectively.
Here, we consider $\beta_{\tau} = \beta_{\nu} = 0.2$. The receive match filter is given by
$w_{rx}(\tau, \nu) = w_{tx}^*(-\tau, -\nu) \, e^{j 2 \pi \nu \tau}$ \cite{Hanly23}.
In Fig.~\ref{fig1} we plot the heat map of $\vert h[k,l] \vert^2$ for a random realization of $h_{\mbox{\scriptsize{phy}}}(\tau, \nu)$ for $\nu_{max} = 1 $ KHz. Notice the delay and Doppler spread of $h[k,l]$ due to which DD carriers interfere and it becomes necessary to perform joint equalization which has high complexity due to lack of structure in the effective DD channel matrix. In Fig.~\ref{fig2} we plot the heat map of the squared magnitude of the DD taps of the precoded channel filter, i.e., $\vert h_a[k,l] \vert^2$ (for the same channel realization in Fig.~\ref{fig1}), which is clearly localized at $(0,0)$ with very little energy leakage.   

Precoding therefore pre-equalizes the channel spread resulting in little interference between DD carriers (i.e., in (\ref{eqn18192}), magnitude of the interference term is significantly smaller than that of the useful term). Hence, \emph{separate equalization} suffices for each DD carrier, i.e., an estimate of the information symbol transmitted
on the $(k,l)$-th DD carrier is given by

{\vspace{-4mm}
\small
\begin{eqnarray}
\label{eqn3232}
    {\widehat x}[k,l] & = & \frac{h_a^*[0,0] \, y_{dd}[k,l] } {\left(\vert h_a[0,0] \vert^2 + \frac{1}{\rho} \right)}.
\end{eqnarray}\normalsize}The equalization complexity of $MN$ information symbols is therefore ${\mathcal O}(MN)$ \emph{only} when compared to ${\mathcal O}(M^3 N^3)$ in the absence of precoding (due to joint equalization which requires inversion of a large $MN \times MN$ matrix). With localized effective precoded channel, receiver only needs to estimate $h_a[0,0]$. We can dedicate a single DD domain carrier to be a pilot carrier and we have no need for guard regions.
The overhead of a single pilot carrier is \emph{significantly smaller} than the overhead without precoding where roughly ${\mathcal O}\left(2 (k_{max} - k_{min}) N \right)$ out of the $MN$ DD carriers are dedicated for channel acquisition.

\begin{table}[!t]
\vspace{-8mm}
    \centering
    \caption{Power-delay profile of Veh-A channel model}
    \vspace{-2mm}
    \begin{tabular}{|c|c|c|c|c|c|c|}
         \hline
         Path index $i$ & 1 & 2 & 3 & 4 & 5 & 6 \\
         \hline
         Delay $\tau_i (\mu s)$ & 0 & 0.31 & 0.71 & 1.09 & 1.73 & 2.51 \\
         \hline
         Relative power (dB) & 0 & -1 & -9 & -10 & -15 & -20 \\
         \hline
    \end{tabular}
    \label{tab1}
    \vspace{-3mm}
\end{table}

For acquiring an estimate of $h_a[0,0]$, we dedicate the $(k_p, l_p) = (M/2, N/2)$-th DD carrier as the pilot carrier with pilot symbol $x[k_p, l_p] = \sqrt{E (MN -1)\eta}$ where
$\eta$ is the pilot to data power ratio (PDR) (pilot energy is $\eta E (MN-1)$ and the energy of all $(MN-1)$ information symbols is $(MN-1)E$). The estimate of $h_a[0,0]$ acquired from $y_{dd}[k_p, l_p]$ is

{\vspace{-4mm}
\small
\begin{eqnarray}
\label{eqn3434}
    {\widehat h_a}[0,0] & \Define & \frac{ y_{dd}[k_p, l_p]}{\sqrt{E (MN -1)\eta} \left( 1 + \frac{1}{\rho \eta (MN -1)}\right)}.
\end{eqnarray}\normalsize}

\section{Numerical Results}
\label{simsec}
We plot the symbol error rate (SER) performance of uncoded $4$-QAM information symbols for the proposed Zak-OTFS precoded system and for conventional Zak-OTFS where joint equalization of all $MN$ DD carriers is performed. For conventional Zak-OTFS we use a dedicated guard and pilot region which is a strip along the Doppler axis having width of $9$ bins along the delay axis (similar to Fig.$3$ in \cite{twobytwopaper}), and estimation is carried out as described in eq. $(30)$ of \cite{zakotfs2}. The channel model and Zak-OTFS parameters are same as that in Fig.~\ref{fig1}. For precoded Zak-OTFS a single pilot is used without any guard region and estimation is using (\ref{eqn3434}). Equalization is carried out using (\ref{eqn3232}) with $h_a[0,0]$ replaced by its estimate. 
\begin{figure}[!t]
\vspace{-8mm}
    \centering
    \includegraphics[width=6.5cm, height=4.1cm]{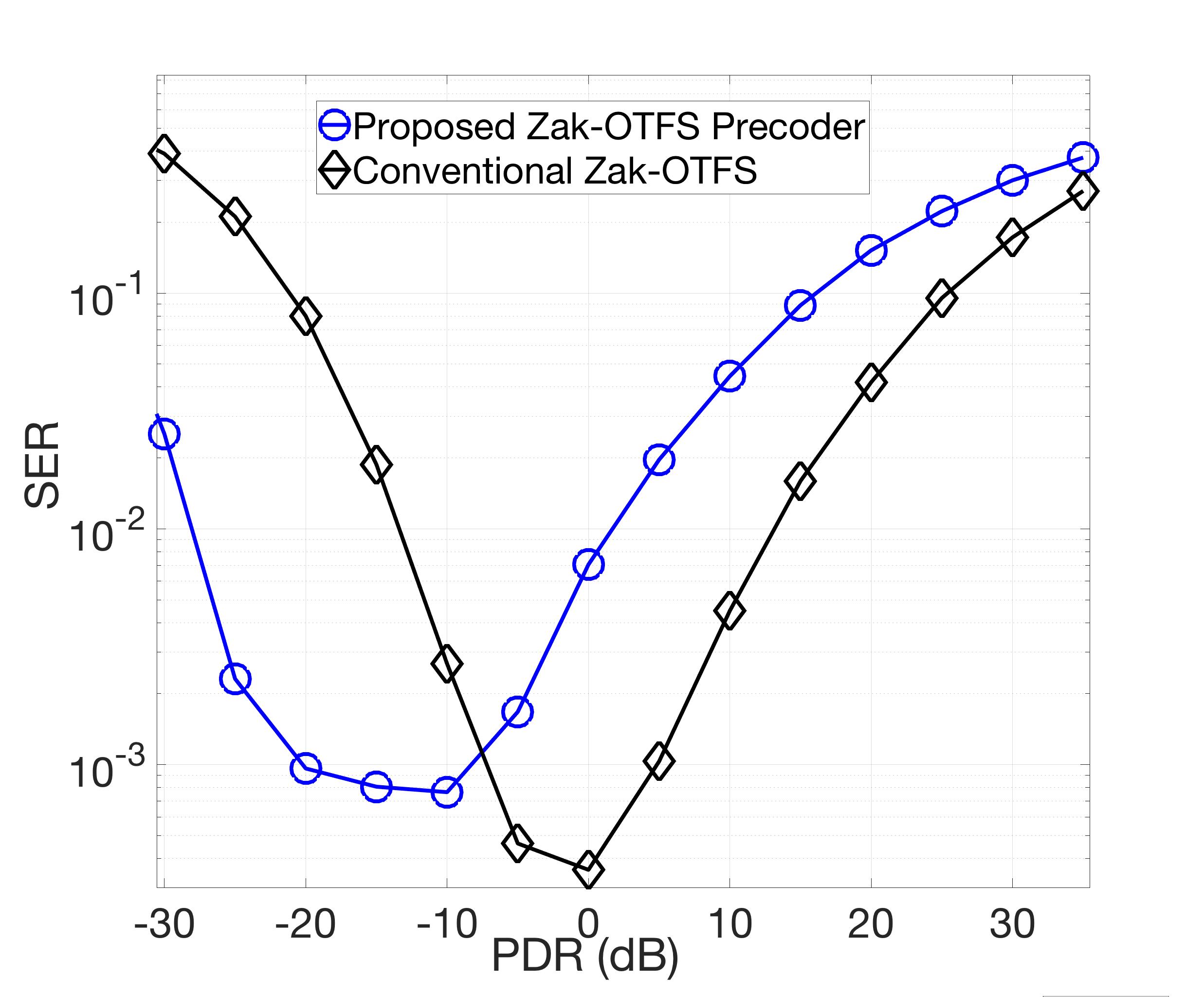}
    \vspace{-4mm}
    \caption{SER vs. PDR ($\eta$). SNR $\rho = 15$ dB, $\nu_{max} = 1$ KHz.}
    \label{fig3}
    \vspace{-4mm}
\end{figure}
\begin{figure}[!t]
    \centering
    \includegraphics[width=7.0cm, height=4.5cm]{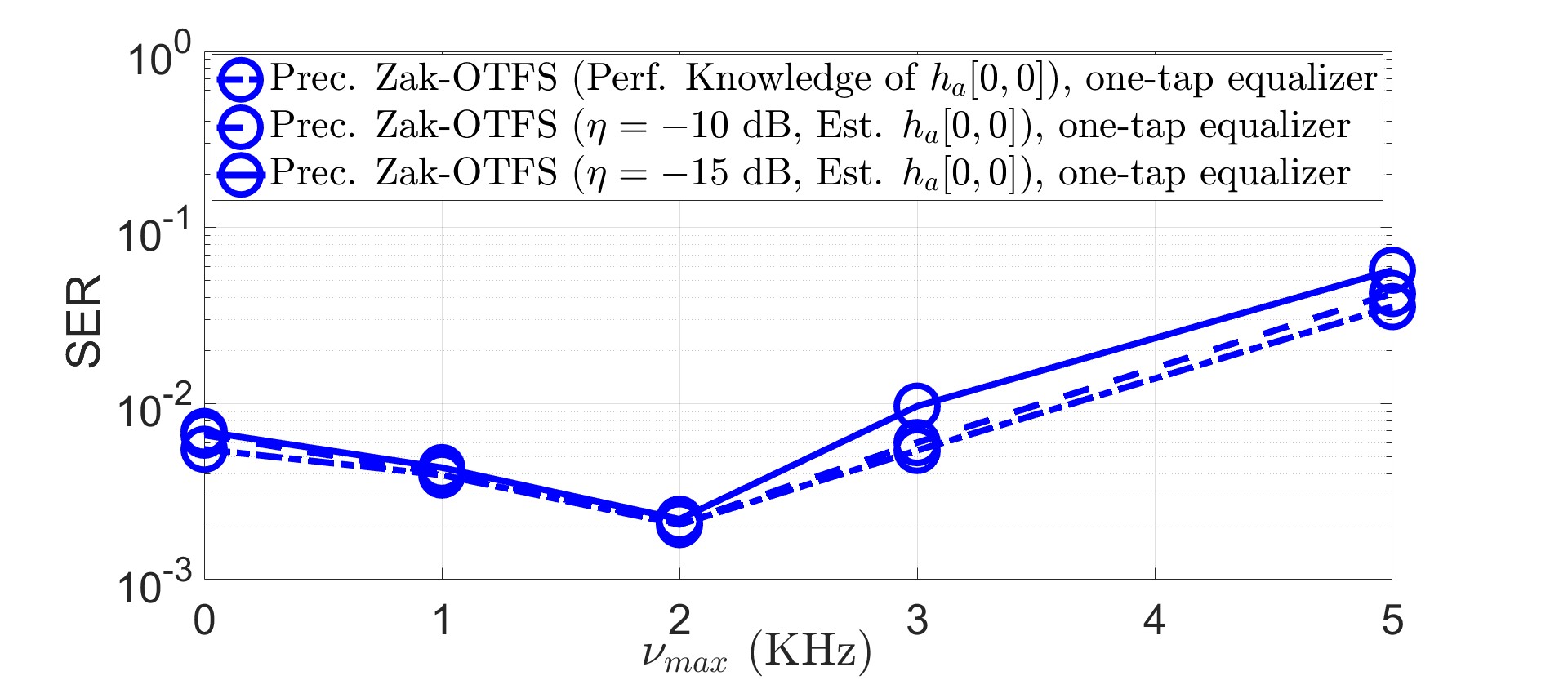}
    \vspace{-4mm}
    \caption{SER vs. $\nu_{max}$ (in KHz).}
    \label{fig4}
    \vspace{-4mm}
\end{figure}

In Fig.~\ref{fig3} we plot SER vs. PDR ($\eta$) for a fixed SNR $\rho = 15$ dB and $\nu_{max} = 1 $ KHz. We observe the characteristic ``U" shaped curve for both systems since initially with increase in $\eta$ the channel estimation accuracy improves resulting in SER improvement but when $\eta$ is high the strong pilot creates interference for data carriers which degrades SER performance. Interestingly, the optimal $\eta$ for precoded Zak-OTFS is only $-10$ dB as compared to $0$ dB for conventional Zak-OTFS. This is because in precoded Zak-OTFS there is little interference between pilot and data carriers since $h_a[k,l]$ is localized, due to which even small pilot power is enough to accurately estimate the I/O relation. This advantage of precoded Zak-OTFS translates to using a larger fraction of the total power for data carriers resulting in higher throughput. 
\begin{figure}[!t]
    \centering
    \includegraphics[width=6.5cm, height=4.1cm]{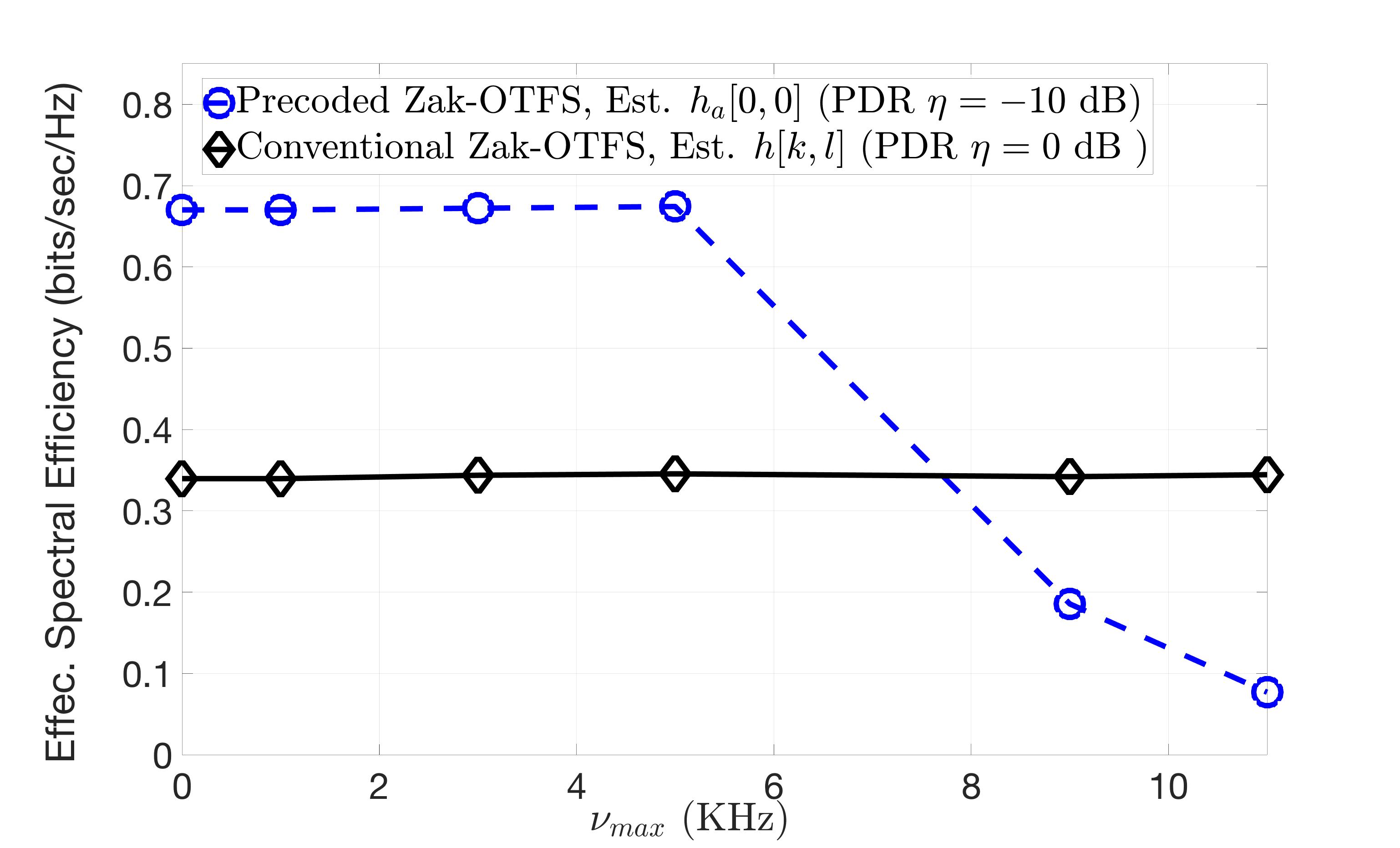}
    \vspace{-4mm}
    \caption{Effective spec. efficiency vs. $\nu_{max}$ (in KHz).}
    \label{fig5}
    \vspace{-4mm}
\end{figure}
In Fig.~\ref{fig4} we plot SER vs. $\nu_{max}$ for a fixed total received power (pilot and data) to noise ratio of $20$ dB.
%Performance of the precoded Zak-OTFS system with one-tap equalizer at the receiver is comparable to that of the conventional Zak-OTFS system even for $\nu_{max} = 2$ KHz (i.e., a high Doppler spread of $4$ KHz). However, for $\nu_{max} > 2$ KHz, the SER performance of precoded Zak-OTFS degrades since $h_a[k,l]$ is no more localized for very high Doppler shifts.
SER performance of precoded Zak-OTFS with estimated ${\widehat h}_a[0,0]$ ($\eta = -10$ dB) is almost same as that with perfect knowledge of $h_a[0,0]$.     

In Fig.~\ref{fig5} we compare the effective spectral efficiency achieved by precoded Zak-OTFS with one-tap equalizer to that achieved by conventional Zak-OTFS with joint equalization, as a function of increasing $\nu_{max}$, for a fixed total received power (pilot and data) to noise ratio of $15.4$ dB. We simulate coded $4$-QAM (with rate half Low Density Parity Check (LDPC) code in 3GPP 5G standard \cite{3gpp}). Spectral efficiency is given by $(1 - \mbox{\scriptsize{BLER}}) N_I / (B T (1 + \beta_{\nu}) (1 + \beta_{\tau}))$, where BLER is the average block error rate and $N_I$ is the number of information bits (before coding) transmitted in each frame. It is observed that even for a very high $\nu_{max} = 5$ KHz (i.e., Doppler spread of $10$ KHz), the spectral efficiency of precoded Zak-OTFS is almost double that of conventional Zak-OTFS. This is due to the large pilot and guard carrier overhead in conventional Zak-OTFS
as compared to only a single DD pilot carrier (with no guard carriers) in precoded Zak-OTFS. Simulation of the TD samples of the precoded transmit signal reveals that its PAPR is same as that of non-precoded Zak-OTFS. 

\section{Conclusions}
%We have developed precoding techniques for the two-dimensional partial response channel defined by twisted convolution that is determined by the action of a doubly spread channel on Zak-OTFS modulation. 
We have described a method of precoding Zak-OTFS carriers in the DD domain that renders the effective I/O relation almost free of interference between carriers.
This enables separate equalization of each DD carrier, which is much less complex than joint equalization of all carriers. We also demonstrate that precoding leads to significant improvements in spectral efficiency. We assume channel state information at the transmitter and making this information available is future work. We also defer exploration of more advanced techniques such as Tomlinson-Harashima precoding to future work.

\end{document}